\documentclass[10pt,aps,pra,twocolumn,nofootinbib]{revtex4-2}
\usepackage[utf8]{inputenc}
\usepackage{graphicx} 
\usepackage{amsmath, amssymb}
\usepackage{xcolor}
\usepackage[normalem]{ulem}
\usepackage{hyperref}
\usepackage{verbatim}
\usepackage{titlesec}
\definecolor{verde}{rgb}{0, 0.56, 0.65}
\usepackage{comment}
\usepackage{braket}
\usepackage{natbib}

\date{April 2025}
\begin{document}
\title{Degrees of Entanglement in Systems of Three Indistinguishable Bosons: Revisiting the Greenberger-Horne-Zeilinger State}

\author{P. Céspedes$^{1,2}$, F. H. Holik$^3$, and A. P. Majtey$^{1,2}$ }

\affiliation{$^1$ Facultad de Matem\'atica, Astronom\'ia, F\'isica y Computaci\'on, Universidad Nacional de C\'ordoba, X5000HUA C\'ordoba, Argentina \\ $^2$ Instituto de F\'isica Enrique Gaviola, CONICET - UNC, C\'ordoba, Argentina\\
$^3$ Instituto de F\'isica La Plata, CONICET - UNLP, 1900 La Plata, Argentina}

\begin{abstract}
    While the concept of entanglement for distinguishable particles is well established, defining entanglement and non-locality in systems of indistinguishable particles, which require the use of the (anti)symmetrization postulate, remains challenging, and multiple approaches have been proposed to address this issue. In this work we study the problem of detecting genuine tripartite entanglement among systems of indistinguishable bosons. A genuine entangled state is one that cannot be separable under any bipartition, where separability in the indistinguishable regime is defined by the existence of single particle properties within each subsystem, without the possibility of knowing which property belongs to which subsystem. We use an algorithm that allows us to search for these single particle properties and, consequently, rank states according to their degree of separability. In particular, we introduce a state of indistinguishable bosons with analogous properties to those of the standard GHZ state.
\end{abstract}

\maketitle

\section{Introduction} 
Indistinguishability is a fundamental property that differentiates quantum mechanics from classical physics, with significant implications for the comprehension and manipulation of quantum systems. Indeed, the peculiar features of systems of identical particles in quantum physics raised deep foundational questions from the very beginning of the theory.

Quantum systems of the same kind can be prepared in states in which their constituents are indistinguishable by any operational means. In other words, exchanging any two particles does not change the composite system's physical properties. As a consequence, in certain circumstances, the attribution of properties to the individual components of a multipartite quantum system cannot be done in a consistent way. This has deep implications for the analysis of non-classical correlations for systems of multiple bosons or fermions \cite{Ghirardi_2002, ghirardi_2004}.

Recent theoretical and experimental advances have demonstrated that indistinguishability can be exploited as a resource for quantum information processing, computation, and metrology \cite{killoran_2014,Morris_2020,lofranco_2018, Mahdavipour2024,Barros2020,cavalcanti2007,Sun2020,Nosrati2020}. A pivotal aspect of this resource is the entanglement that arises from the exchange symmetry of identical particles. A number of approaches have been put forth with the aim of extracting useful or accessible entanglement from systems whose correlations originate only on the symmetrization of the wave function \cite{killoran_2014,bouvrie_2017_b,martin2024}. Notable developments include the formulation of a resource theory for fermionic systems \cite{gigena_2020,gigena_2021,Morris_2020}, experimental evidence of entanglement generation in ultracold gases through Bose-Einstein Condensate (BEC) splitting \cite{Lange_2018, Vitagliano2023}, and theoretical studies on the generation of entanglement between molecular condensates \cite{bouvrie_2019}. These developments indicate the broad applicability of indistinguishability as a resource across different areas of quantum information science.

Notwithstanding, the study and treatment of the non-classical correlations originated in systems of identical particles still remains challenging, driving a variety of approaches to address these issues
\cite{Ghirardi_2002,ghirardi_2004,Eckert_2002,Dowling2006, shi2003, Paskauskas2001, plastino_2009_epl, Zhou_2009, Li2001, compagno2016, debarba2019}. 

In particular, while there exist several efficient tools for dealing with fermionic systems \cite{Eckert_2002,tichy_2011_JPB,plastino_2009_epl,Majtey2023,Benatti_2020,ding2022,Ding2024}, there are still many open questions for the boson case, with the mode entanglement approach \cite{zanardi_2000} (distinct from particle entanglement) being by far the most extensively studied in these systems \cite{Dalton_Identical_Bosons_I,Dalton_Identical_Bosons_II}. Regarding the particle approach, one of the ways in which the separability of systems of indistinguishable particles can be determined is by analysing their properties as represented by projection operators. The key observation is that for the distinguishable case a pure state can be considered separable if and only if it is possible to define a complete set of properties for each of its constituents. A similar criterion can be used for the indistinguishable case, but keeping in mind that it is no longer possible to determine which particle has which property. For a comprehensive review of the properties-based analysis in systems of indistinguishable particles, we refer the reader to \cite{Ghirardi_2002,tichy_2011_JPB,Benatti_2020}. 

In this work we use the properties-based approach to analyze particle entanglement in systems of indistinguishable bosons. We go beyond previous works by studying the entanglement of systems of three indistinguishable qubits and qutrits. Our work shows that qudit states allow for a larger variety of degrees of entanglement, showing different features than the case of qubit states. Our analysis allows for a classification of the states of these systems in different classes, according to how much separable (or non-separable) they are. We pay special attention to the case of GHZ-like states, and show that these are quite different when we jump from two to three particles. Our method has the advantage that it can be implemented numerically, allowing for the characterization of entanglement for higher dimensional models (in both the number of components, and their dimensionality). 

Our results may have implications for a deeper understanding and characterization of devices dealing with indistinguishable bosons. In the last decade, the idea of building quantum devices based on photonic systems led to a promising area of research \cite{Slussarenko-2019,Takeda-Furusawa}. There are already programmable photonic quantum computers available in the market in which indistinguishability plays a key role \cite{Maring2024}. Remarkably, systems of indistinguishable bosons were used in experiments aimed to demonstrate quantum advantage \cite{Madsen2022}. Our work can help to understand and quantify the resources produced by such technologies.

The paper is organized as follows. In section \ref{s:Properties_and_entanglement_review} we review how to connect the correlations of the compound system with the definiteness of its constituents properties for distinguishable and indistinguishable particles. In section \ref{s:Three_Particle_Case} we analyze the three-indistinguishable-bosons qubits and qudits cases. We discuss some particular examples of GHZ and W-like states. Finally, in section \ref{s:Conclusions} we show our conclusions.

\section{Separability criterion}\label{s:Properties_and_entanglement_review}

\subsection{Distinguishable particles} \label{s:distinguishable_particles}

In order to categorize quantum states in a systematic manner according to their level of entanglement, it is first necessary to establish a formal definition of this property. In this context, we focus on a quantum state comprising two distinguishable $d$-dimensional subsystems, which are commonly referred to as a bipartite system of qudits. The more complex case involving entanglement among more than two particles will be addressed later.

A bipartite quantum state \(|\psi\rangle \in \mathcal{H}_1 \otimes \mathcal{H}_2\) is called separable if it can be written as a product state, i.e., if one can find single-particle states \(|\phi_i\rangle \in \mathcal{H}_i\) such that

\begin{equation}
    \label{eq:puredissep}
|\psi\rangle = |\phi_1\rangle \otimes |\phi_2\rangle. 
\end{equation}
This state is completely determined by the single-particle states \(|\phi_1\rangle\) and \(|\phi_2\rangle\). The outcomes of measurements on a pure separable state are therefore not correlated at all: they are fully defined by the pure states of the subsystems. Indeed, we can simply state that the first particle is prepared in \(|\phi_1\rangle\), while the second particle is prepared in \(|\phi_2\rangle\), which corresponds to the assignment of definite physical properties to the components of the system.  

In what follows, and in full agreement with the finding of a realistic description of the system's constituents \cite{epr1935}, we will define separability in terms of the existence of a complete set of properties for the subsystems. This will provide us with a useful criterion to rigorously define entanglement in systems of identical particles. Thus, a quantum state is considered not entangled \cite{epr1935} if a complete set of properties can be attributed to each individual subsystem \cite{Ghirardi_2002}, meaning that a projective measurement can be designed for each subsystem in such a way that its outcome can be predicted with certainty. In other words, there exists an observable such that measuring it on the given non-entangled state reveals the `pre-existing' values of the system, similar to what occurs in classical mechanics. This corresponds to finding a realistic description of the system's components \cite{epr1935}. If such a realistic description exists, no Bell-type inequality can be violated.

Formulating the previous notion mathematically, it can be stated that in a pure state \(|\psi\rangle\) describing a composite two-distinguishable-particle system, the subsystem \(S_1\) is non-entangled with subsystem \(S_2\) if there exists a one-dimensional projection operator \(\hat{P}\) (with eigenvalue 1, by definition) such that
\begin{equation} \label{eq:proyector_dist}
\langle \psi | \hat{P}_1 \otimes \mathbb{I}_2 | \psi \rangle = 1.
\end{equation}

If this condition holds, we can assert that the first subsystem possesses the physical properties defined by the operator \(\hat{P}\) \textcolor{red}{\cite{ghirardi_2004}}. The outcome of measuring \(\hat{P}\) is certain, indicating that the first subsystem was prepared in the unique eigenstate of \(\hat{P}\).

The physical notion of `possessing a complete set of properties' is equivalent to the formal notion of separability defined in Eqn. \eqref{eq:puredissep}. For any separable pure state, its properties are naturally described by projections onto the quantum states of its constituent particles. However, if no product state decomposition for a given state exists, such well-defined constituent properties can no longer be identified. Thus, the scenario changes for entangled states that cannot be expressed as direct products (like (\ref{eq:puredissep})) of two single-particle states. The information contained within an entangled state is not fully determined solely by the states of the individual subsystems. As an example\textcolor{blue}{,} let us consider the entangled state given by:
\begin{equation}
| \psi \rangle = \frac{1}{\sqrt{2}} (| 0 1 \rangle - | 1 0 \rangle),
\end{equation}
which represents a coherent superposition of a two-body state where the first particle is in the internal state \( | 0 \rangle \) and the second particle is in the state \( | 1 \rangle \), and vice versa, with the first particle in \( | 1 \rangle \) and the second one in \( | 0 \rangle \).

The outcomes of a projective measurement performed on the first subsystem in the (orthonormal) basis \( \{ | 0 \rangle , | 1 \rangle \} \) are completely undetermined, with each possible outcome occurring with equal probability \( \frac{1}{2} \): 

\begin{eqnarray}
p(0) &=& \langle \psi | (| 0 \rangle \langle 0 | \otimes \mathbb{I}_2) | \psi \rangle = \langle \psi | (| 1 \rangle \langle 1 | \otimes \mathbb{I}_2) | \psi \rangle\nonumber\\ &=& p(1) = \frac{1}{2}.
\end{eqnarray}

The same is true for a similar measurement performed on the second subsystem, where no prediction can be made for individual outcomes. However, the combined measurement results on the two subsystems are perfectly anti-correlated: once a measurement result \( j = 0 \) is obtained on the first subsystem, the two-body state is projected onto the state \( | 0 1 \rangle \), allowing one to predict with certainty that the measurement result of the other subsystem will be \( 1 \).

These definitions and arguments can be straightforwardly extended to states composed by more than two particles. In multi-particle states, it is usual to identify different types of entanglement \cite{amico_2008, dur2000, Guhne_2010}. For example, in the three-partite case, we have:

\begin{itemize}
    \item totally separable states (A-B-C), $|\psi\rangle=|a\rangle\otimes|b\rangle\otimes|c\rangle$. In this case, each particle has a well defined and compete set of properties.
    \item biseparable states (A-BC, B-CA, C-AB). A state $| \psi \rangle_{ABC} $ is biseparable if it can be written as $| \psi \rangle_{A} \otimes | \psi \rangle_{BC} $ under some bipartition, where $|\psi \rangle_{BC}$ might be a two-particle entangled state.
    \item totally entangled, which means that there exist no separable bipartition. This family includes W-like and GHZ-like entangled states that determine two inequivalent classes of entanglement (they cannot be connected by local operations and classical communication).
\end{itemize}

Next, we extend these definitions for the case of indistinguishable particles.

\subsection{Two indistinguishable particles}

As already stated, throughout this paper we will adhere to the definition of entanglement according to which pure states that are incompatible with the existence of a set of well-defined properties of  all the individual subsystems will be regarded as entangled, or non-separable. Equivalently,  separable, or non-entangled states, are those for which a set of well-defined properties exists for every particle (see \cite{Majtey2023} for an introduction on this subject).

Mathematically, and in accordance to (\ref{eq:proyector_dist}), we will assert that one of the particles (but we do not know which one) of the pure indistinguishable-particle state $\rho = \ket{\psi} \bra{\psi}$ has a complete set of properties described by the one-dimensional projection operator $P$ if 

\begin{equation} \label{eq:projector_2particle}
Tr[\rho \mathcal{E}_P ] = 1, \text{ } \mathcal{E}_P = P \otimes (\mathbb{I} - P) + (\mathbb{I} - P) \otimes P.
\end{equation}

Symmetrized observables, such as those in Eq. \eqref{eq:projector_2particle}, become relevant in scenarios involving compound systems of identical bosons (or fermions), where correlations arising from the permutational symmetry of the state cannot be neglected. In such cases, it is impossible to consistently assign labels to the particles, and after measurement, one cannot determine which particle is which. Even in the absence of interactions, the permutational symmetry of the state vector can give rise to correlations that have measurable effects. If non-symmetric projections were used in these situations (as in the case of distinguishable particles discussed in Section \ref{s:distinguishable_particles})), one might erroneously conclude that entanglement is present, even if the particles had never interacted. The use of symmetrized operators allows us to correctly distinguish between correlations resulting from genuine entanglement and those arising solely from permutational symmetry.

The symmetric projector $\mathcal{E}_P$ can be interpreted as follows: either one of the particles has the set of properties associated to $P$ and the other does not, or vice versa. With this definition, the condition $Tr[\rho \mathcal{E}_P ] = 1$ implies that one of the particles has the property $P$, but we do not know which one \cite{Benatti_2014, ghirardi_2004}. 

For the bosons case, the state will be regarded as separable if there exists another projector $Q$, either orthogonal or equal to $P$, satisfying $Tr[\rho \mathcal{E}_Q ] = 1$, i.e, both constituents possess a complete set of properties, but it cannot be asserted which one has which property. If this were the case, the properties would render the particles distinguishable \cite{Ghirardi_2002, Benatti_2014}. 
In contrast to the situation that arises in the case of fermions, in which the possibility of ascribing a comprehensive set of properties to one subgroup necessarily implies that the remaining subgroup possesses an orthogonal set of properties, this is no longer the case for systems of identical bosons, since it might happen that a non-orthogonal set of properties can be attributed to the other subgroup. For a thorough analysis of the two-particle fermionic and bosonic case, we refer the reader to \cite{Ghirardi_2002}, \cite{ghirardi_2004}, \cite{Benatti_2014}.

\section{Three indistinguishable particles}\label{s:Three_Particle_Case}

After establishing how to determine quantum properties and their relation to entanglement and separability in both the distinguishable and indistinguishable regimes, we will now apply these concepts to the case of three indistinguishable bosons states.

In analogy to (\ref{eq:projector_2particle}), we look for a projector operator that will allow us to represent the fact that one of the three particles of the pure bosonic state $\ket{\psi}$ might have a property $P$ \footnote{In this work we will restrict to rank one properties.}, without being possible to establish which particle has it. 

To this effect we define three new projectors:

\begin{subequations}
\begin{align}
    \mathcal{E}_{P_1} &= P \otimes (I-P) \otimes (I-P) + (I-P) \otimes P \otimes (I-P) \notag \\ \label{epsilon_p1}
    &\quad + (I-P) \otimes (I-P) \otimes P \\
    \mathcal{E}_{P_2} &= P \otimes P \otimes (I-P) + P \otimes (I-P) \otimes P \notag \\
    &\quad + (I-P) \otimes P \otimes P  \\
    \mathcal{E}_{P_3} &= P \otimes P \otimes P, 
\end{align}
\end{subequations}

\noindent where $\langle \psi| \mathcal{E}_{P_i} |\psi  \rangle = 1$, $ i =1,2,3$ gives the probability that \textit{exactly}   one, two or the three particles of the system possess the property $P$ \cite{ghirardi_2004}. It is important to notice that the terms $(I-P)$ imply that, for example, in $\mathcal{E}_{P_1}$, one particle has $P$ with certainty, and the others do not have $P$ with certainty (though we do not know which one). Thus, $\mathcal{E}_{P_1}$ represents the property ``one particle has $P$ with certainty, and the others do not have $P$ with certainty". 

Since the above three projectors are orthogonal with each other, we can form a new projector by summing them:

\begin{equation}
    \mathcal{E}_{P} = \mathcal{E}_{P_1} + \mathcal{E}_{P_2} + \mathcal{E}_{P_3},
\end{equation}

\noindent where $\langle \psi| \mathcal{E}_{P} |\psi \rangle$ gives the probability of finding \textit{at least} one of the particles of the state $|\psi \rangle$ in the state onto which $P$ projects, $\ket{\psi}$ being a symmetric superposition of states where the particles may or may not have the property $P$. Therefore, the condition $\langle \psi| \mathcal{E}_P |\psi \rangle = 1$ implies that \textit{at least} one particle has the complete set of properties described by the single particle projector $P$. In the same sense, we could also think of a projector with the form $\mathcal{E}_{P_2} + \mathcal{E}_{P_3}$ meaning that at least \textit{two} particles possess the property $P$.

From the definition in \cite{Ghirardi_2002}, it follows that for the case of three particles, the identical constituents $S_1$, $S_2$ and $S_3$ of a composite quantum system $S = S_1 +S_2 + S_3$ are separable when each of the three constituents possess a complete set of properties. 

Consequently, in order to find whether a given system is separable, we have to answer the following questions: does any of the particles possess the property $P$? If the answer is yes, does any of the particles possess the property $Q$, either equal or orthogonal to $P$? If the answer is again yes, does the last particle possess a property $R$, either equal or orthogonal to $P$ and $Q$? If it does, the state is separable. In case that a set of properties satisfying these conditions cannot be found, the state must be regarded as entangled. Thus, the algorithm for determining whether a state is separable or not (and if not, to which degree), can be summarized as follows:

\begin{enumerate}
\item Given an arbitrary $|\psi\rangle$ check whether there exists $P=|\phi_0\rangle\langle\phi_0|$ such that $\langle\psi| \mathcal{E}_P|\psi\rangle=1$. Notice that this equation can be solved numerically, using a suitable parametrization of the one dimensional projection $P$. If such $P$ does not exist, the state must be considered as completely entangled. If it does, follow the next steps.
\item Similarly, look for a one dimensional projection $Q$ such that it is either equal or orthogonal to $P$, satisfying $\langle\psi| \mathcal{E}_Q|\psi\rangle=1$. If such a $Q$ exists, follow the next step. If it does not, the state can be considered as partially entangled.
\item Finally, look for a one dimensional projection $R$ such that it is either equal or orthognal to $P$ and $Q$, satisfying $\langle\psi| \mathcal{E}_R|\psi\rangle=1$. If such a $R$ exists, the state can be considered as completely separable. If this is not the case, the state can be regarded as partially entangled, although to a lesser degree than in the preceding step.
\end{enumerate}

\noindent It should be noted that the aforementioned algorithm can be extended to an arbitrary number of particles in a manner consistent with the original formulation.

We now focus on a system of three indistinguishable bosons. We start with a general state where it can be stated beforehand that at least one of the particles has the complete set of one particle properties $P=| \phi_0 \rangle \langle \phi_0 |$, where $ \{| \phi_i \rangle\}_{i \geq 0}$, $i = 0 ... d-1$, is an orthonormal basis in the single particle Hilbert space whose first element is the state corresponding to the mentioned property. We are interested in finding whether such a state also has properties $Q$ and $R$ either equal to or orthogonal to $P$.

In this basis, a general three particle state can be written as:

\begin{equation}
    | \psi \rangle =  \sum_{i,j,k=0} C_{ijk} | \phi_i \rangle | \phi_j \rangle | \phi_k \rangle,
\end{equation}

\noindent with $C_{ijk}$ symmetric under permutations of the indices. After explicitly separating the first element of the basis, the state takes the form:

\begin{equation} \label{general_state_total}
\begin{split}
    | \psi \rangle & =   C_{000} | \phi_0 \rangle | \phi_0 \rangle | \phi_0 \rangle 
    + \sum_{i \neq 0} C_{i00} | \phi_i \rangle | \phi_0 \rangle | \phi_0 \rangle 
    \\ 
    & + \sum_{j \neq 0} C_{0j0} | \phi_0 \rangle | \phi_j \rangle | \phi_0 \rangle
     +  \sum_{k \neq 0} C_{00k} | \phi_0 \rangle | \phi_0 \rangle | \phi_k \rangle 
    \\
    & + \sum_{i,j \neq 0} C_{ij0} | \phi_i \rangle | \phi_j \rangle | \phi_0 \rangle
    +  \sum_{j,k \neq 0} C_{0jk} | \phi_0 \rangle | \phi_j \rangle | \phi_k \rangle
    \\
     & + \sum_{i,k \neq 0} C_{i0k} | \phi_i \rangle | \phi_0 \rangle | \phi_k \rangle,
\end{split}
\end{equation}

\noindent where we have used that the condition $ \langle \psi| \mathcal{E}_P |\psi \rangle = 1 $ implies that $C_{ijk} = 0 $ $ \forall i,j,k \neq 0$. Furthermore, using that the coefficients are symmetric under permutations, we obtain:

\begin{equation} \label{general_state}
\begin{split}    
    | \psi \rangle & = C_{000} | \phi_0 \rangle | \phi_0 \rangle | \phi_0 \rangle 
    \\ & +  \sum_{i \neq 0} C_{i00} (| \phi_i \rangle | \phi_0 \rangle | \phi_0 \rangle + | \phi_0 \rangle | \phi_i \rangle | \phi_0 \rangle + | \phi_0 \rangle | \phi_0 \rangle | \phi_i \rangle )
    \\ & + \sum_{i,j \neq 0} C_{ij0} (| \phi_i \rangle | \phi_j \rangle | \phi_0 \rangle + | \phi_i \rangle | \phi_0\rangle | \phi_j \rangle + | \phi_0 \rangle | \phi_i \rangle | \phi_j \rangle ).
\end{split}
\end{equation}

\noindent Notice that normalization implies that $ |C_{000}|^2 + 3 \sum_{i,j \neq 0} |C_{ij0}|^2 + 3 \sum_{i \neq 0} |C_{i00}|^2 =1 $.

The state $| \psi \rangle$ will be separable, in case that we can find one dimensional projection operators $Q$ and $R$, either orthogonal or equal to $P$, for which $\langle \psi| \mathcal{E}_Q |\psi \rangle = 1$ and $\langle \psi| \mathcal{E}_R |\psi \rangle = 1$. Whenever these conditions are fulfilled, it is possible to declare that each one of the three particles possesses a complete set of properties, but without telling which one has which \cite{Ghirardi_2002}. It follows that the mere fact that the state of a system composed of identical particles is symmetric (or antisymmetric), does not necessarily preclude the attribution of a complete set of physical properties to its subsystems. If the state can be expressed as the symmetrization of orthogonal (or the same, in the case of bosons) states, it will not be entangled. Notably, we show below that there exist three particle systems that can be written as a symmetrization of states with well-defined properties, but not orthogonal among each other (see, for example, Eqn. \eqref{eq:cero_cero_theta} below). In this case, even if properties of individual systems are well defined, the state cannot be considered separable. 

\subsection{Three indistinguishable qubits}

We now analyze the case in which each particle behaves as a qubit (for example, photons having two orthogonal polarizations modes). Let us use the same basis $ \{ \ket{\phi_0},\ket{\phi_1} \}$ for each particle, where $\ket{\phi_0}$ and $\ket{\phi_1}$ are fixed but otherwise arbitrary orthonormal vectors. The first thing we notice is that there cannot exist three mutually orthogonal properties $P$, $Q$ and $R$, due to the fact that this is not allowed for the case of two-dimensional single-particle Hilbert spaces. If a state happens to be separable, it will mean that two or more particles share the same one-particle state. A general three-particle state of qubit systems satisfying the condition $ \bra{\psi} \mathcal{E}_P \ket{\psi}=1$ as in (\ref{general_state}), acquires the form: 

\begin{equation} \label{general_state_01}
\begin{split}
    | \psi \rangle & =  C_{000} | \phi_0 \rangle | \phi_0 \rangle | \phi_0 \rangle 
    \\
      & +  C_{100} (| \phi_1 \rangle | \phi_0 \rangle | \phi_0 \rangle 
    + | \phi_0 \rangle | \phi_1 \rangle | \phi_0 \rangle
    + | \phi_0 \rangle | \phi_0 \rangle | \phi_1 \rangle )
    \\
     &  + C_{110} (| \phi_1 \rangle | \phi_1 \rangle | \phi_0 \rangle
    +  | \phi_0 \rangle | \phi_1 \rangle | \phi_1 \rangle
    + | \phi_1 \rangle | \phi_0 \rangle | \phi_1 \rangle ),
\end{split}
\end{equation}
since it is a superposition of states where exactly one, two and three particles have the property  $P=\ket{\phi_0}\bra{\phi_0}$.

In what follows we present a series of illustrative examples. We shall start with the state W.
\begin{equation} \label{eq:w_state}
    | \psi \rangle = \frac{| \phi_1 \rangle | \phi_0 \rangle | \phi_0 \rangle
    +  | \phi_0 \rangle | \phi_1 \rangle | \phi_0 \rangle
    + | \phi_0 \rangle | \phi_0 \rangle | \phi_1 \rangle}{\sqrt{3}}.
\end{equation}

\noindent By writing the symmetrized projections $\mathcal{E}_{P_1}$, $\mathcal{E}_{P_2}$, $\mathcal{E}_{P_3}$ and $\mathcal{E}_{P}$ in terms of $P=\ket{\phi_0}\bra{\phi_0}$, a straightforward calculation yields:

\begin{equation}
\begin{split}
\langle \psi| \mathcal{E}_{P_1} |\psi \rangle &= 0
\\
\langle \psi| \mathcal{E}_{P_2} |\psi \rangle &= 1
\\
\langle \psi| \mathcal{E}_{P_3} |\psi \rangle &= 0
\\
\langle \psi| \mathcal{E}_{P} |\psi \rangle &= 1.
\end{split}
\end{equation}

\noindent Similarly, for $Q= | \phi_1 \rangle \langle \phi_1 |$, we have: 

\begin{equation}
\begin{split}
\langle \psi| \mathcal{E}_{Q_1} |\psi \rangle &= 1
\\
\langle \psi| \mathcal{E}_{Q_2} |\psi \rangle &= 0
\\
\langle \psi|\mathcal{E}_{Q_3} |\psi \rangle &= 0
\\
\langle \psi|\mathcal{E}_{Q} |\psi \rangle &= 1.
\end{split}
\end{equation}

Recalling the meaning of each projector, it follows that exactly two particles have the property $P$ and exactly one particle has the property $Q$. Since it is possible to attribute a complete set of properties to each particle, the state must be considered separable, in a striking difference with regard to the usual W state of the distinguishable particles case. 

Let us now consider the state:

\begin{equation} \label{c110=0}
\begin{split}
    | \psi \rangle & =  C_{000} | \phi_0 \rangle | \phi_0 \rangle | \phi_0 \rangle 
    \\
   & +  C_{100} (| \phi_1 \rangle | \phi_0 \rangle | \phi_0 \rangle
    +  | \phi_0 \rangle | \phi_1 \rangle | \phi_0 \rangle
    + | \phi_0 \rangle | \phi_0 \rangle | \phi_1 \rangle )
\end{split}
\end{equation}

\noindent with $C_{000}\neq 0$ and $C_{100}\neq 0$. A straightforward calculation shows:

\begin{equation}\label{e:ProjectionP}
\begin{split}
\langle \psi| \mathcal{E}_{P_1} |\psi \rangle &= 0
\\
\langle \psi| \mathcal{E}_{P_2} |\psi \rangle &= 3 C_{100}^2
\\
\langle \psi| \mathcal{E}_{P_3} |\psi \rangle &= C_{000}^2
\\
\langle \psi| \mathcal{E}_{P} |\psi \rangle &= 1.
\end{split}
\end{equation}

\noindent Similarly, we can check that for $Q= | \phi_1 \rangle \langle \phi_1 |$, we have: 

\begin{equation}\label{e:ProjectionQ}
\begin{split}
\langle \psi| \mathcal{E}_{Q_1} |\psi \rangle &= 3 C_{100}^2
\\
\langle \psi| \mathcal{E}_{Q_2} |\psi \rangle &=0
\\
\langle \psi| \mathcal{E}_{Q_3} |\psi \rangle &= 0
\\
\langle \psi| \mathcal{E}_{Q} |\psi \rangle &= 3 C_{100}^2 \neq 1.
\end{split}
\end{equation}

Using Eq. \eqref{e:ProjectionP}, it follows that at least one particle has the property $P$. Even more, using normalization, we have that $\langle \psi| \mathcal{E}_{P_2}+\mathcal{E}_{P_3} |\psi \rangle = 1$. Thus, we can assert that at least two particles have the property $P$. On the other hand, Eqn. \eqref{e:ProjectionQ} implies that the remaining particle does not have the property $Q$ (which is orthogonal to $P$). Given that in two dimensions $Q$ is the only candidate for an orthogonal property to $P$, it follows that there exists no orthogonal property for the remaining particle. Therefore, this state must be regarded as an entangled one.

The above result becomes clearer if we write the state as a symmetrization of the product of non-orthogonal states:

\begin{equation} \label{eq:cero_cero_theta}
    |\psi\rangle = N ( |\Theta\rangle |\phi_0\rangle |\phi_0\rangle 
    + |\phi_0\rangle |\Theta\rangle |\phi_0\rangle + |\phi_0\rangle  |\phi_0\rangle |\Theta\rangle )
\end{equation}

\noindent with
\begin{equation*}
    |\Theta\rangle = \frac{1}{N} \left( \frac{C_{000}}{3} |\phi_0\rangle + C_{100} |\phi_1\rangle \right)  
\end{equation*}
and
\begin{equation*}
N=\sqrt{ \left( \frac{C_{000}}{3} \right)^2 + C_{100}^2}.
\end{equation*}

\noindent Since $\bra{ \phi_0}\Theta \rangle \neq 0$, the above expression shows explicitly that two of the particles have the property $|\phi_0\rangle\langle\phi_0|$, and the remaining one has the property $|\Theta\rangle\langle\Theta|$, but these are not orthogonal.

Sometimes we might be dealing with states where we do not know beforehand if any property $P$ exists, i.e states which might include the coefficient $C_{111}$ in the expression (\ref{general_state_01}). Therefore, for a general state, we first need to identify if at least one particle has a property. If such a property does not exists, the state can be considered entangled. If it does, one can rewrite the state in the corresponding basis and proceed similarly as before, i.e., keep looking for properties for the remaining particles. In order to illustrate this methodology, consider the following state: 

\begin{equation} \label{eq:ghz_qubit}
    | \psi \rangle = \frac{|0 0 0 \rangle + |1 1 1 \rangle}{\sqrt{2}}
\end{equation}

At first sight, it appears that the state is totally entangled, as in the case of distinguishable particles, since it is not immediately apparent whether any of the particles possess a complete set of properties and it has the form of the standard distinguishable GHZ state, $| GHZ \rangle = \frac{|000 \rangle + |111  \rangle}{\sqrt{2}}$. However, the behaviour of this state in the indistinguishable regime is different. Moreover, as we will show, the three particle case (the state of Eqn. \eqref{eq:ghz_qubit}) is also different than the two particle case (i.e., $| \psi \rangle = \frac{|0 0 \rangle + |1 1  \rangle}{\sqrt{2}}$). In reference \cite{ghirardi_2004}, the authors demonstrate the existence of one-particle properties for the latter. Upon writing it in the corresponding symmetrized form, it is proven that the state is, in fact, totally separable.
However, in the three-particle case, using a numerical implementation of the steps $1-3$ described in Section \ref{s:Three_Particle_Case} we find that at least one particle in (\ref{eq:ghz_qubit}) possesses the property $P = \ket{\phi_0} \bra{\phi_0}$, with

$$ | \phi_0 \rangle = \frac{1}{\sqrt{2}} | 0 \rangle + \frac{1}{\sqrt{2}} | 1 \rangle . $$

Choosing a new basis 
 $$ \{ | \phi_0 \rangle , | \phi_1 \rangle \} = \{ \frac{1}{\sqrt{2}} | 0 \rangle + \frac{1}{\sqrt{2}} | 1 \rangle , \frac{1}{\sqrt{2}} | 0 \rangle - \frac{1}{\sqrt{2}} | 1 \rangle \} $$ 

\noindent the state takes a form such as in (\ref{general_state_01}):

\begin{equation}
\begin{split}
    \ket{\psi} & =  \frac{1}{\sqrt{4}}  \ket{\phi_0}\ket{\phi_0}\ket{\phi_0} 
    \\ 
    & + \frac{1}{\sqrt{4}}  (\ket{\phi_1}\ket{\phi_1}\ket{\phi_0} + \ket{\phi_1}\ket{\phi_0}\ket{\phi_1} + \ket{\phi_0}\ket{\phi_1}\ket{\phi_1} ).
\end{split}
\end{equation}

\noindent We can rewrite this expression in the symmetrized form

\begin{equation} \label{eq:ghz_sym}
\begin{split}
    | \psi \rangle = \frac{1}{3} ( |\phi_0\rangle |\Phi\rangle |\Theta\rangle + |\phi_0\rangle |\Theta\rangle |\Phi\rangle + |\Phi\rangle |\Theta\rangle |\phi_0\rangle 
    \\
    + |\Phi\rangle |\phi_0\rangle |\Theta\rangle + |\Theta\rangle |\Phi\rangle |\phi_0\rangle + |\Theta\rangle |\phi_0\rangle |\Phi\rangle )   
\end{split}  
\end{equation}

with 
$$\Theta = \frac{-i}{\sqrt{4}}|\phi_0 \rangle + \frac{\sqrt{3}}{\sqrt{4}} |\phi_1 \rangle $$ 

and 

$$ \Phi = \frac{i}{\sqrt{4}}|\phi_0 \rangle + \frac{\sqrt{3}}{\sqrt{4}} |\phi_1 \rangle.$$

We can now calculate the desired expectation values $\langle \psi |\mathcal{E}_{P_i} | \psi \rangle$, $i=1,2,3$, with $P = | \phi_0 \rangle \langle \phi_0 |$ and proceed with the same analysis as before. 
A straightforward calculation shows that at least one particle has the property $P$, but it is not possible to find other properties orthogonal or equal to $P$. Since $ \langle \phi_0|\Theta\rangle \neq 0$, $ \langle \phi_0|\Phi\rangle \neq 0$ and $ \langle \Theta|\Phi\rangle \neq 0$, even if the state (\ref{eq:ghz_qubit}) can be written as a symmetrization of single-particle states, it remains entangled, in contrast to the two-particle case. It is also not maximally entangled, as opposed to the distinguishable GHZ state. In section \ref{s:three_qutrits} we propose, for the indistinguishable boson case, a state with characteristics analogous of those of the standard GHZ state.

As a last example, consider the following state:

\begin{equation}\label{eq:General_Superposition}
\begin{split}
| \psi \rangle = \frac{1}{\sqrt{8}} ( | 0  0  0 \rangle +      | 1  0  0 \rangle 
    + | 0  1  0 \rangle
    + | 0  0  1 \rangle
    \\
    + | 1  1   0 \rangle
    +  | 0  1  1 \rangle
    + |  1 0  1 \rangle + | 1  1  1 \rangle )
\end{split}
\end{equation}

By evaluating $\bra{\psi} \mathcal{E}_P \ket{\psi}$ we find that the three particles possess the property $P= \ket{\phi_0} \bra{\phi_0}$, with

\begin{equation}
    | \phi_0 \rangle = \frac{1}{\sqrt{2}} | 0 \rangle + \frac{1}{\sqrt{2}} | 1 \rangle .
\end{equation}

Indeed, the state can be written as a symmetrical product state: 

\begin{equation}
    | \psi \rangle = | \phi_0 \rangle | \phi_0 \rangle | \phi_0 \rangle  
\end{equation}

Thus, the state turns out to be separable. The reason can be intuitively understood using the circuit model of quantum computing. Such state can be easily produced in a quantum computer by applying Hadamard gates to each qubit of the $|000\rangle$ state: $|\psi\rangle=H\otimes H \otimes H|000\rangle$. This means that it can be produced by applying local unitaries to a completely separable state, and then, it cannot contain any non-classical correlations (even in the distinguishable particles scenario).

\subsection{Three indistinguishable qutrits}\label{s:three_qutrits} 

As already stated, when looking for orthogonal properties $P$, $Q$ and $R$ in three indistinguishable qubit states, at least two of the particles must have the same property in order that the state be separable. However, in three qutrit states, the higher dimensionality makes room for accessing three distinct orthonormal properties, and new possibilities take place. In this situation, we want to determine whether the entanglement of the system follows some comparable classification to the distinguishable particle case as mentioned in \ref{s:Properties_and_entanglement_review}. 

When considering indistinguishable particles, as stated before, separability is no longer defined in terms of product states, but in terms of particles possessing properties. With this definition, a state would be biseparable if we can assign a complete set of properties to one of the particles, leaving the other two in a possibly entangled two-particle state, i.e. a state as described by \cite{Ghirardi_2002}. In addition, the particle of the separable bipartition must possess a property that is orthogonal to the properties of the remaining part. If these two conditions hold, the system can be separated in two parts, being the properties of one part orthognal to those of the other.

As an example, consider the state (\ref{general_state}):

\begin{equation}
\begin{split}
      | \psi \rangle = C_{110} (| \phi_0 \rangle | \phi_1 \rangle | \phi_1 \rangle
    +  | \phi_1 \rangle | \phi_0 \rangle | \phi_1 \rangle
    + | \phi_1 \rangle | \phi_1 \rangle | \phi_0 \rangle ) 
    \\
     + C_{012} (| \phi_0 \rangle | \phi_1 \rangle | \phi_2 \rangle
    +  | \phi_0 \rangle | \phi_2 \rangle | \phi_1 \rangle
    + | \phi_1 \rangle | \phi_0 \rangle | \phi_2 \rangle
    \\
    + | \phi_1 \rangle | \phi_2 \rangle | \phi_0 \rangle 
     + | \phi_2 \rangle | \phi_1 \rangle | \phi_0 \rangle 
      + | \phi_2 \rangle | \phi_0 \rangle | \phi_1 \rangle )
\end{split}
\end{equation}

\noindent which can be rewritten as the symmetrization of the product state $|\phi_0\rangle |\phi_1\rangle |\Theta\rangle$:

\begin{equation} \label{eq:qutrit_bi_sep}
\begin{split}
    | \psi \rangle =  N & ( |\phi_0\rangle |\phi_1\rangle |\Theta\rangle + |\phi_0\rangle |\Theta\rangle |\phi_1\rangle + |\phi_1\rangle |\phi_0\rangle  |\Theta\rangle
     \\ + & |\phi_1\rangle  |\Theta\rangle |\phi_0\rangle + |\Theta\rangle |\phi_0\rangle |\phi_1\rangle + |\Theta\rangle |\phi_1\rangle |\phi_0\rangle )   
\end{split}  
\end{equation}
\normalsize

\noindent with 
$$ \Theta = \frac{1}{N} \left( \frac{C_{011}}{2} |\phi_1 \rangle + C_{012} |\phi_2 \rangle \right)$$
and
$$N=\sqrt{ \left( \frac{C_{011}}{2} \right)^2+C_{012}^2}. $$

It is clear that $\langle \phi_0 | \Theta \rangle = 0$ and $\langle \phi_1 | \Theta \rangle, \langle \phi_2 | \Theta \rangle \neq 0$. If we consider the property $P=|\phi_0 \rangle \langle \phi_0|$, we can write the corresponding projector $\mathcal{E}_{P_1}$ to find:

\begin{equation}
\langle \psi| \mathcal{E}_{P_1} |\psi \rangle = 3 C_{011}^2 + 6 C_{012}^2 = 1
\end{equation}

\noindent i.e exactly one particle has the property $P$ (notice that this implies that at least one has $P$). 

After determining that exactly one particle has the property $P$, one concludes that it is certain that the other two do not have the property $P$. Moreover, from equation \eqref{epsilon_p1}, it can be inferred that if the other two particles have defined properties, their properties must be orthogonal to $P$. In other words, the remaining two particles have properties that are contained in a space orthogonal to the one defined by $P$.  It then follows by a straightforward calculation that at least one of the other two particles possesses a complete set of properties $Q=|\phi_1 \rangle \langle \phi_1|$ and at least another one has the property $R=|\Theta \rangle \langle \Theta|$ which are not orthogonal between each other. Thus, the subsystem that possesses the properties $Q$ and $R$ must be considered to be in an entangled state. 

In light of the aforementioned considerations, we can conclude that this state can be regarded as biseparable, in analogy to the distinguishable particle case, in the sense that one subsystem which consists of a particle with property $P$ is not entangled with the other subsystem, which consists of two entangled particles. 

As an example of a totally separable three-qutrit state, consider:

\begin{equation} \label{eq:qutrit_separable}
\begin{split}
    \ket{\psi}  =& \frac{1}{\sqrt{6}}  ( \ket{\phi_0} \ket{\phi_1} \ket{\phi_2} + \ket{\phi_0} \ket{\phi_2} \ket{\phi_1} + \ket{\phi_1} \ket{\phi_0} \ket{\phi_2}
    \\
    + & \ket{\phi_0} \ket{\phi_2} \ket{\phi_1} + \ket{\phi_2} \ket{\phi_1} \ket{\phi_0} + \ket{\phi_2} \ket{\phi_0} \ket{\phi_1})
\end{split}
\end{equation}

\noindent which is the symmetrization of a product state of three orthogonal single-particle states. It should be clear that in this case, the particles have the properties $P=\ket{\phi_0}\bra{\phi_0}$, $Q=\ket{\phi_1}\bra{\phi_1}$ and $R=\ket{\phi_2}\bra{\phi_2}$, but we cannot tell which property corresponds to each particle.

\subsubsection{GHZ-like states for systems of three qutrits}

Finally, consider the following state,  which we propose as an adequate GHZ state for indistinguishable particles:

\begin{equation} \label{eq:ghz_3}
    \ket{\psi} = \frac{\ket{000} + \ket{111} + \ket{222}}{\sqrt{3}}. 
\end{equation}

There is no projector operator such that $\bra{\psi} \mathcal{E}_X \ket{\psi} = 1$ for any property $X$. 
Therefore we can affirm that this state shows genuine tripartite entanglement. Moreover, the reduced density matrix for any of the particles reads:

\begin{equation}
    \rho = \frac{1}{3} ( \ket{00}\bra{00} + \ket{11}\bra{11} + \ket{22}\bra{22} ).
\end{equation}

The above equation means that randomly choosing any two bosons gives place to a separable state \cite{ghirardi_2004}, which strengthens the analogy to the distinguishable GHZ state.

Working with qutrits allowed us to generalise the GHZ state for indistinguishable particles. Can we do the same for the W state? If we consider the generalised W state for qutrits \cite{Kim_2008},

\begin{equation}
    \ket{\psi} = \frac{1}{\sqrt{6}} (\ket{100} + \ket{010} + \ket{001} + \ket{200} + \ket{020} + \ket{002}),
\end{equation}

\noindent we notice that it can be written in the symmetrized form

\begin{equation} \label{eq:w_qutrit}
    \ket{\psi} = \frac{1}{\sqrt{3}} (\ket{\Theta} \ket{0} \ket{0} + \ket{0} \ket{\Theta} \ket{0} + \ket{0} \ket{0} \ket{\Theta} 
\end{equation}
with

\begin{equation}
    \ket{\Theta} = \frac{\ket{1}+\ket{2}}{\sqrt{2}}. 
\end{equation}

\noindent Two of the particles possess the property $P=\ket{0}\bra{0}$, while the remaining one possesses the property $Q=\ket{\Theta} \bra{\Theta}$. Since $P$ is orthogonal to $Q$, the state remains separable and is not conceptually different from the qubit case \ref{eq:w_state}. This shows that increasing the dimension is ineffective in achieving the objective of generalizing the W state for indistinguishable bosons, as was the case with the GHZ state. Notably, for systems of identical fermions, analogous states for both the W and GHZ states exist. It was shown in \cite{martin2024} that these states require a single-particle Hilbert space of dimension 6.

After the exhaustive analysis of these distinct situations we can comprehensively categorise every type of state, according to their properties: 

\begin{itemize}
    \item States where the three particles have a complete set of properties, either equal or orthogonal among each other [$P \perp Q$, $P \perp R$, $Q \perp R$]. These are all separable states, corresponding to the examples studied in Eqns. \eqref{eq:w_state},\eqref{eq:General_Superposition},\eqref{eq:qutrit_separable} and \eqref{eq:w_qutrit};

    \item States where one particle has a complete set of properties orthogonal to both of the properties of the other two particles, which in turn are not orthogonal among them [$P \perp Q$, $P \perp R$, $Q \not\perp R$]. These are biseparable states, with the state of Eqn. \eqref{eq:qutrit_bi_sep} being an example;

    \item States where two particles have a complete set of properties, either equal or orthogonal, while the state of the third one is not orthogonal to any of those [$P \perp Q$, $P \not\perp R$, $Q \not\perp R $], This is a partially entangled state, exemplified by Eqn. \eqref{eq:cero_cero_theta};

    \item States with three properties that are not orthogonal among each other [$P \not\perp Q$, $P \not\perp R$, $Q \not\perp R $]. This corresponds to a partially entangled state, exemplified by that of Eqn. \eqref{eq:ghz_sym};

    \item States with no properties at all. This case corresponds to completely entangled states, such as the one given by Eqn. \eqref{eq:ghz_3}.
\end{itemize}

The above classification indicates a rich structure of degrees of non-separability for the three qutrits case and extends the results of previous works.

\section{Conclusions}\label{s:Conclusions}

In this work we studied the non-classical correlations of three-particle bosonic states of qubits and qutrits. By using an operational approach based on the analysis of the properties associated to the system's constituents, we were able to determine different degrees of separability. Our results explicitly show that there is a richer variety of separability cases for three qudits than for two or three qubits. Remarkably, our method can be implemented numerically, and thus, extended in principle to higher dimensional systems and states for which no analytic solution can be found. 

It is notable that the standard GHZ state, when considered as the state of a system of three indistinguishable qubits, turns out to be not-fully entangled. We recovered a state with analogous properties to those of the standard GHZ state of distinguishable particles, by introducing GHZ like states of three indistinguishable qutrits \eqref{eq:ghz_3}. These states maintain the features of being maximally non-separable, in the sense that no non-trivial property can be defined for the system's constituents, and of leaving the remaining two particles in a separable state after tracing over any of the three particles, as is the case for the standard GHZ state of distinguishable particles. 

Our findings verify that non-classical correlations associated with systems of indistinguishable particles are considerably more involved than those observed in distinguishable systems. In particular, bosons exhibit a more complex scenario than fermions, and a systematic approach is still missing. The classification of states according to their separability properties introduced in this work calls for the development of entanglement measures tailored for three-qudit bosonic states. 
For instance, one might claim that for systems of indistinguishable bosons, at least three particles with a single-particle Hilbert space of dimension greater than two are required in order to recover states with properties analogous to those of the distinguishable qubits GHZ state. 
A similar phenomenon occurs for identical fermions. In this case, the smallest system exhibiting entanglement is a system comprising two subsystems with a Hilbert space of dimension four and the fermionic GHZ state is obtained for three fermions with a single-particle Hilbert space dimension of six \cite{Eckert_2002,majtey_2016, martin2024}. 
Whether these assertions holds with higher number of particles is an open question which will be addressed in future works.  
\vspace{-0.20cm}
\begin{acknowledgments}
\vspace{-0.25cm}
P.C. and A.P.M. acknowledge funding from Grants No. PICT 2020-SERIEA-00959 from ANPCyT (Argentina) and No. PIP 11220210100963CO from CONICET (Argentina) and partial support from SeCyT, Universidad Nacional de Córdoba (UNC). F.H.H. acknowledges funding from Grants No. PICT-2019-2019-01272 from ANPCyT (Argentina). A.P.M. is grateful to Andrea Valdés-Hernández for fruitful discussions.
\end{acknowledgments}

\end{document}